\documentclass[12pt]{article}

\usepackage{amssymb,amsmath,mathrsfs,tikz,graphicx}
\def\S{\Sigma}

\def\be{\begin{equation}}
\def\ee{\end{equation}}
\def\bea{\begin{eqnarray}}
\def\eea{\end{eqnarray}}
\def\bean{\begin{eqnarray*}}
\def\eean{\end{eqnarray*}}

\begin{document}

\title{Junction conditions for $F(R)$-gravity and their consequences}
\author{Jos\'e M. M. Senovilla \\
F\'{\i}sica Te\'orica, Universidad del Pa\'{\i}s Vasco, \\
Apartado 644, 48080 Bilbao, Spain \\ 
josemm.senovilla@ehu.es}
\date{}
\maketitle
\begin{abstract} 
I present the junction conditions for $F(R)$ theories of gravity and their implications: the generalized Israel conditions and equations. These junction conditions are necessary to construct global models of stars, galaxies, etc., where a vacuum region surrounds a finite body in equilibrium, as well as to describe shells of matter and braneworlds, and they are stricter than in General Relativity in both cases. For the latter case, I obtain the field equations for the energy-momentum tensor on the shell/brane, and they turn out to be, remarkably, the same as in General Relativity. An exceptional case for quadratic $F(R)$, previously overlooked in the literature, is shown to arise allowing for a discontinuous $R$, and leading to an energy-momentum content on the shell with unexpected properties, such as non-vanishing components normal to the shell and a new term resembling classical dipole distributions. For the former case, they do not only require the agreement of the first and second fundamental forms on both sides of the matching hypersurface, but also that the scalar curvature $R$ and its first derivative $\nabla R$ agree there too. I argue that,
as a consequence, {\em matched} solutions in General Relativity are not solutions of $F(R)$-models generically. Several relevant examples are analyzed.
\end{abstract} 


\section{Introduction}
Alternative theories of gravity, such as models based on an $F(R)$ Lagrangian, have received considerable attention in the past years, see e.g. \cite{SF,CF,NO}. The basic fundamental solutions of standard General Relativity (GR), such as the Schwarzschild and Kottler spherically symmetric exteriors, or the Friedmann-Lema\^{\i}tre-Robertson-Walker cosmological models, are also solutions of the $F(R)$-theories. In the latter case, however, the energy-momentum content of the space-time differs from that of GR. Concerning the former case, and as an illustrative example of the greater richness of $F(R)$ theories, the Birkhoff-Jensen theorem on the uniqueness of the exterior (vacuum) solution of a spherically symmetric gravitating system no longer holds and thus there arise many other (even non-static) vacuum solutions different from the Schwarzschild solution \cite{CF}. Similarly, many possible interior solutions (be they perfect fluids or more general matter) are feasible in the extended theories.
It is tacitly understood, in general, that the GR solutions are a very good approximation to the $F(R)$ solutions at least in some regimes such as low energy ones \cite{CF}.

In this paper, the question of whether or not {\em global models} in GR describing (say) stars in equilibrium or collapsing bodies that form black holes ---in both cases all the way from the centre up to infinity---remain as viable valid models in the generalized $F(R)$ theories will be addressed. The related problem of inserting cavities into standard cosmological models, which requires a similar analysis, has been recently considered in \cite{CDGN}. The conclusions obtained herein agree with theirs. In both cases, one needs the proper junction conditions through the hypersurface separating the exterior and the interior of the global gravitational field, in particular, through the world-surface of finite bodies such as stars. These junction conditions have been obtained in \cite{DSS} via a calculation performed in Gaussian coordinates relative to the matching hypersurface, and were later used in \cite{CDGN}. The calculation in \cite{DSS} allows for cases with branes or thin shells at the matching hypersurface. However, the field equations for the energy-momentum tensor on the brane (formulas (\ref{new1}) and (\ref{new2}) below) were not derived in \cite{DSS},  and they do not seem to have been written down hitherto. Remarkably, they are actually identical with those of GR, even though the derivation as well as the expression for the energy-momentum tensor are quite different in $F(R)$ theories. This leads to some speculations about its possible universal character.

To avoid loading the paper with heavy calculations, I derive the correct junction conditions for $F(R)$ gravity in an Appendix by using tensor distributions \cite{L,T}. The main results are succinctly presented in section \ref{sec:matchingconditions}. For non-linear functions $F(R)$, the junction conditions {\em always} require continuity of the trace of the second fundamental form of the matching hypersurface as well as, {\em generically}, the continuity of the scalar curvature $R$. It turns out that matter shells and braneworlds cannot be umbilical hypersurfaces, and the brane tension is proportional to the discontinuity of the derivative of $R$, in contrast with the GR result, see however \cite{BD,DSS} and references therein.

Very surprisingly, there also arises an exceptional case which seems to have been overlooked in the literature and where a {\em discontinuous} scalar curvature $R$ is allowed. This exceptional case can only arise in theories with a quadratic function $F(R)$,  and leads to quite a different matter content on the brane or thin shell so that, for instance, components normal to the matching hypersurface can arise as well as a completely unexpected new term, described by (\ref{t}) and (\ref{Omega}) below, which resembles those of classical ``dipole distributions'' ---a kind of Dirac-delta-prime distribution. The interpretation of such a new term is quite unclear. This is considered in subsection \ref{subsec:exceptional} where the general field equations for the energy-momentum content of the shell in this exceptional situation are presented. 

The case without braneworlds or thin shells, so that the curvature tensor distributions do not possess singular parts, is then considered in subsection \ref{subsec:normal}
and their implications in terms of the energy-momentum tensor quantities ---equation (\ref{Isrnew}) below--- are derived. This seems to be also new and again adopts the very same form as in GR, despite this not being trivial at all. This allows to prove, for example, the important result that the matching hypersurface for a compact perfect fluid is always defined by the vanishing of the pressure, and then the fluid has also vanishing energy density there or at least it becomes tangent to the matching hypersurface ---see subsection \ref{subsec:apps}.

The junction conditions turn out to be more restrictive that in the GR case, as they will impose ---in addition to the same conditions as in GR--- differentiability of the scalar curvature $R$ across the matching hypersurface. 
This will lead to a simple but general proof that, in general, GR matched solutions will not be solutions of generalized $F(R)$ gravity theories, as analyzed in section \ref{sec:discussion}. In particular, for example, the Oppenheimer-Snyder collapsing star \cite{OS} to form a black hole is not a solution of any $F(R)$-theory. The corresponding complementary matching in the sense of \cite{FST}, which describes the Einstein-Straus vacuole \cite{ES}, see \cite{MMV}, is also impossible for non-linear $F(R)$ as recently demonstrated in \cite{CDGN}. The constant-density interior Schwarzschild solution matched to its exterior is not a solution either, nor are the vast majority of the multiple static spherically symmetric solutions with a perfect fluid interior matched to the exterior vacuum Schwarzschild solution ---see also, in this respect, the discussion in \cite{ESK,HS}. I also consider more general cases with dynamical and radiating exteriors, and the conclusion is the same: GR matched solutions are not solutions of the extended $F(R)$ theories generically. A very particular model that is a global solution of both GR and $F(R)$ theories is found, however, given by a Robertson-Walker interior with radiation equation of state matched to the Vaidya radiating exterior solution \cite{V}. The complementary matching gives rise to a radiating vacuole in an expanding universe.

I must mention that there have been several papers discussing the different problems that arise concerning exact solutions of $F(R)$ theories, such as \cite{F,BSM,KM} and many references therein, where curvature singularities appear or there arise impediments to have strong gravitational fields in spherically symmetric stars.
The difficulties shown herein are, nevertheless, of an intrinsic different nature as I am just concerned with the question of whether or not {\em matched GR} solutions can be solutions of the extended theories.

\section{The matching conditions}
\label{sec:matchingconditions}
The appropriate framework to study the matching of two different spacetimes across a timelike hypersurface\footnote{The whole study can be performed for null hypersurfaces, and in general for hypersurfaces changing their causal character from point to point \cite{MS}. However, to keep the presentation as simple as possible and make the main point plain I have preferred to restrict myself to the important case of a timelike matching hypersurface.} is that of tensor distributions \cite{L,T}, because the proper junction conditions follow from analyzing the singular parts of some curvature tensor distributions and, for the proper case with no brane/shell, by demanding that they vanish. This leads in GR to the standard Darmois and Israel matching conditions \cite{D,I} requiring the agreement, on the matching hypersurface, of the first fundamental form, while supplying a formula (respectively, field equations) for the brane/shell energy-momentum content in terms of the discontinuities of the second fundamental forms (resp.\ of the energy-momentum tensors) inherited from both sides of the space-time . In particular, there is no shell/brane when the second fundamental forms agree, in which case the junction conditions are equivalent, in a certain sense, to the Lichnerowicz $C^{1}$ conditions on the metric components {\em in admissible coordinates} \cite{L1,D,I}. 
For a summary of the junction conditions and standard references see section 3.8 in \cite{Exact} and \cite{MS}. I present a derivation of the junction conditions following this general method in the Appendix, and I only give in this section a summary of the results, stressing the new formulas that were not found before and the appearance of a particular, extraordinary case, for theories with a quadratic function $F(R)$.

Following the notation used in the Appendix, let $\Sigma$ be the timelike matching hypersurface, $n_\mu$ the unit normal to $\S$, $h_{\mu\nu}=g_{\mu\nu}-n_\mu n_\nu $ its first fundamental form and $K_{\mu\nu}$ its second fundamental form. This last object, as well as others, may have a jump across $\S$. For any function $f$ its jump across $\S$ is denoted by $[f]$.

\subsection{Case allowing for matter shells or branes}
It is shown in the Appendix that in general $F(R)$ theories the following condition must hold
\be
\left[K^\mu{}_\mu\right] =0 \, .\label{Kcont}
\ee
Therefore, the trace of the second fundamental form must always be continuous across $\Sigma$, even for cases with shells of matter or branes. This is quite different from the GR case, and forbids the use of umbilical hypersurfaces ---characterized by $K_{\mu\nu}=f h_{\mu\nu}$--- one of the most common cases to describe braneworlds in GR. 

Now the analysis splits into two possibilities, depending on whether or not $F'''(R)=0$.
\subsubsection{The generic case: $F'''(R)\neq 0$}
In this case, as shown rigorously in the Appendix, the requirement
\be
 \left[R\right]=0 \label{Rcont}
\ee
is unavoidable. Hence, the scalar curvature must always be continuous across $\Sigma$, even for cases with shells of matter or branes.

The energy-momentum tensor in the brane or thins shell reads (see Appendix)
\be
\kappa \tau_{\mu\nu}= -F'(R_\Sigma) \left[K_{\mu\nu}\right] + F''(R_\Sigma) n^\rho \left[\nabla_{\rho}R\right] h_{\mu\nu},
\hspace{1cm} n^\mu \tau_{\mu\nu} =0. 
\label{newIsr}
\ee
Thus, the following result holds:
\begin{quotation}
The proper junction conditions allowing for shells of matter or branes in $F(R)$ theories with $F'''(R)\neq 0$ are the agreement of the first fundamental forms on both sides of the matching hypersurface together with (\ref{Rcont}) and (\ref{Kcont}). The energy-momentum content of the shell or brane is then given by formula (\ref{newIsr}).
\end{quotation}
It should be noted that the first summand in (\ref{newIsr}) is traceless due to (\ref{Kcont}), so that the trace of the energy-momentum singular part (sometimes called the ``brane tension'') reads simply
$$
\kappa\, \tau^\rho{}_{\rho} =(n-1) F''(R_\Sigma)\,  n^\rho \left[\nabla_{\rho}R\right]
$$
and is fully determined by the discontinuity of the normal derivative of $R$ across $\Sigma$. This is in sharp contrast with the GR case, where the brane tension is the discontinuity of $K^\rho{}_\rho$, which is always continuous now. 
In particular, for $\Sigma$ with {\em only} brane tension $\lambda$ so that $\kappa \tau_{\mu\nu} = \lambda h_{\mu\nu}$ one must have 
$[K_{\mu\nu}]=0$ and the tension is $\lambda =F''(R_\Sigma)\,  n^\rho \left[\nabla_{\rho}R\right]$.

Concerning the equations satisfied by the energy-momentum tensor in the brane, one has from the Appendix
\bea
(K^+_{\rho\sigma}+K^-_{\rho\sigma})\tau^{\rho\sigma} = 2 n^\beta n^\mu \left[ T_{\beta\mu}\right] \label{new1}
\\
\overline\nabla^\beta \tau_{\beta\mu}=-n^\rho h^\sigma{}_\mu \left[ T_{\rho\sigma}\right]. \label{new2}
\eea
As far as I am aware, these relations are new and, most remarkably, identical with the GR case ---corresponding to (\ref{1}) and (\ref{2}) via the Einstein equations. It arises the idea that they may be universally valid for diffeomorphism-invariant theories , but I do not know of any result along these lines.

\subsubsection{The exceptional case: $F'''(R)=0$}
\label{subsec:exceptional}
As proven in the Appendix, this case allows for a discontinuous $R$, so that $[R]\neq 0$ is possible and
 the energy-momentum tensor distribution acquires a singular part with two terms
$$
\underline T_{\mu\nu}=T^+_{\mu\nu} \underline\theta +T^-_{\mu\nu} (1-\underline\theta)+\left(\tau_{\mu\nu}+\tau_\mu n_\nu +\tau_\nu n_\mu +\tau n_\mu n_\nu \right) \underline\delta^{\Sigma} + \underline{t}_{\mu\nu}
$$
where 
\bea
\kappa \tau_{\mu\nu} =-\{1+\alpha(R^++R^-)\}[K_{\mu\nu}] 
+\alpha \left\{2ah_{\mu\nu}-[R](K^+_{\mu\nu}+ K^-_{\mu\nu})\right\}, \, \, n^\mu\tau_{\mu\nu}=0, \label{tauexc}\\
\kappa \tau _\mu =-2\alpha \overline\nabla_\mu [R], \quad \quad n^\mu \tau_\mu =0, \hspace{4cm} \label{tauex}\\
\kappa \tau = 2\alpha [R] K^\rho{}_\rho , \label{taue} \hspace{4cm}\\
\kappa \underline t_{\mu\nu}=2\alpha \underline\Omega_{\mu\nu} .\hspace{4cm} \label{t}
\eea
Here 
$$
\alpha\equiv \frac{1}{2} F''(R)
$$
is a constant, $a$ is a function on $\S$ defined in (\ref{DRdisc2}) and
the delta-prime--type distribution $\underline \Omega$ is defined in (\ref{Omega}). 

Thus, we now have:
\begin{quotation}
The proper junction conditions allowing for shells of matter or branes in $F(R)$ theories with $F'''(R)= 0$ are the agreement of the first fundamental forms on both sides of the matching hypersurface together with (\ref{Kcont}). A discontinuous $R$ is permitted and the energy-momentum content of the shell or brane is given by formulas (\ref{tauexc}-\ref{t}).
\end{quotation}

The discontinuity of the energy-momentum tensor can be computed from (\ref{fe}) with (\ref{Fquadratic}), leading easily to
$$
\kappa [T_{\mu\nu}]=(1+2\alpha R_{\S})[G_{\mu\nu}]+2\alpha\left\{[R]R^{\S}_{\mu\nu}+g_{\mu\nu}[\nabla^{\rho}\nabla_{\rho}R]-[\nabla_{\mu}\nabla_{\nu}R]\right\}
$$
which on using (\ref{D2R2}), (\ref{1}), (\ref{2}) and (\ref{HEi}) provides in this case
\bean
\kappa n^{\mu}n^{\nu}[T_{\mu\nu}]&=& -(1+2\alpha R_{\S})K^{\rho\sigma}_{\S}[K_{\rho\sigma}]+2\alpha \left\{[R]R^{\S}_{\mu\nu}n^{\mu}n^{\nu}+aK^{\rho}{}_{\rho} +\overline\nabla^{\rho}\overline\nabla_{\rho}[R] \right\},\\
\kappa n^{\nu}h^{\rho}{}_{\mu}[T_{\rho\nu}]&=&(1+2\alpha R_{\S})\overline\nabla^{\rho}[K_{\rho\mu}]\\
&+&2\alpha\left\{[R]n^{\nu}h^{\rho}{}_{\mu}R_{\rho\nu}-\overline\nabla_{\mu} a +\left[K_{\rho\mu}\right]\overline\nabla^{\rho}R_{\Sigma}+K^{\S}_{\rho\mu}\overline\nabla^{\rho}[R]\right\}.
\eean
Combining these with (\ref{tauexc}--\ref{taue}) and (\ref{gauss}-\ref{coda}) one obtains
\bean
n^{\nu}h^{\rho}{}_{\mu}[T_{\rho\nu}]&=&-\overline\nabla^{\nu}\tau_{\mu\nu}-K^{\rho}{}_{\rho}\tau_{\mu}-\overline\nabla_{\mu}\tau ,\\
n^{\mu}n^{\nu}[T_{\mu\nu}]-\tau_{\mu\nu}K^{\mu\nu}_{\S}+\overline\nabla^{\mu}\tau_{\mu}&=&2\frac{\alpha}{\kappa} [R]\left(R^{\S}_{\mu\nu}n^{\mu}n^{\nu}+K^{\mu\nu}_{\S}K_{\S\mu\nu} \right)\\
&=&\frac{\alpha}{\kappa} [R]\left(R_{\S}-{\cal R} +(K^{\rho}{}_{\rho})^{2}+K^{+}_{\mu\nu}K^{-\mu\nu}\right).
\eean
It may be observed that the case $\alpha=0$ is simply GR. If on the other hand one sets $[R]=0$ then the generic case is recovered under the assumption (\ref{Fquadratic}). 

The appearance of the last term (\ref{t}) is most remarkable, and very surprising. The classical interpretation of such terms (for instance in electromagnetism) is that they describe a distribution of dipoles on $\S$. This is usually thought to be physically not viable in Gravitation due to its attractive character and the positivity of masses. The interpretation of such new terms in this exceptional case is thus open and may give rise to new possibilities for describing quite exotic braneworld scenarios.

\subsection{Case without shells or branes}
\label{subsec:normal}
Consider now  the proper matching case where no shells of matter or branes are allowed, so that $T_{\mu\nu}$ can have, at most, jump discontinuities. From the Appendix we know that in this situation (\ref{Rcont})  must hold and also
$$
[K_{\mu\nu}]=0, \quad \quad n^\rho \left[\nabla_{\rho}R\right]=0
$$ 
thus
\be
\left[\nabla_{\mu}R\right]=0. \label{DRcont}
\ee
This immediately informs us that the GR junction conditions are necessary, that is to say, the first and second fundamental forms must agree on $\Sigma$ from both sides, {\em however}, in contrast to GR, they are not enough in general as they have to be complemented with (\ref{Rcont}) and (\ref{DRcont}). 
Hence, the following result has been obtained:
\begin{quotation}
The proper junction conditions allowing for no shells of matter nor branes in  $F(R)$ theories with $F''(R)\neq 0$ are those of GR ---the agreement of the first and second fundamental forms on both sides of the matching hypersurface--- together with (\ref{Rcont}) and (\ref{DRcont}).
\end{quotation}
Observe that (\ref{Rcont}) and (\ref{DRcont}) amount to saying that $R$, as a function, is differentiable everywhere (also across $\Sigma$).

Let me consider the implications of these junction conditions. To start with, it should be noted that, in addition to (\ref{Isr}) one also has (\ref{Rcont}), so that 
\be
n^{\mu}\left[R_{\mu\nu}\right]=0, \hspace{1cm} \left[R\right] =0 \label{Isr2}
\ee
and $R_{\mu\nu}$ has $n(n-1)/2-1$ allowed independent discontinuities in contrast to the $n(n-1)/2$ of standard GR. One wonders about the allowed discontinuities for the energy-momentum tensor. They follow from computing the discontinuity of the field equations (\ref{fe}) which, on using (\ref{D2R}) with $a=0$ and $[K_{\mu\nu}]=0$, produce
\be
F'(R_\Sigma)\left[R_{\mu\nu}\right]+A\, h_{\mu\nu} F''(R_\Sigma)=\kappa \left[T_{\mu\nu}\right]  \label{new}
\ee
where $A$ is a function on $\Sigma$ given by (see Appendix)
$$
A\equiv n^\mu n^\nu \left[\nabla_{\mu}\nabla_{\nu}R\right]
$$
that is, $A$ represents the discontinuity in the second normal derivative of $R$ across $\Sigma$.
Taking traces
$$
(n-1) AF''(R_\Sigma) =\kappa \left[T^{\mu}{}_{\mu}\right] \, .
$$
From (\ref{Isr2}) and (\ref{new}) ---alternatively from (\ref{new2}) and (\ref{new1})--- one derives
\be
n^{\mu}\left[T_{\mu\nu}\right]=0  \label{Isrnew}
\ee
which happen to be identical with those in the GR case, proving the continuity of the $n$ normal components of the energy-momentum tensor across $\Sigma$. Nevertheless, in contrast to GR, the $n(n-1)/2$ allowed independent discontinuities for $T_{\mu\nu}$ are not given by those of the Ricci tensor exclusively (as it has one less), but also by the new one encoded in the second derivative of $R$ represented by $A$. In any case, (\ref{Isrnew}), as well as (\ref{new2}) and (\ref{new1}), look like very good properties of $F(R)$ theories.

\subsubsection{Application: perfect fluids}
\label{subsec:apps}
Consider for example the situation where a perfect fluid interior ($V^{-}$) is to be matched to a vacuum exterior solution ($V^{+}$). Then $T^{+}_{\mu\nu}=0$ while
\be
T^-_{\mu\nu}=\tilde\varrho \tilde u_\mu \tilde u_\nu +\tilde p (g_{\mu\nu}+\tilde u_\mu \tilde u_\nu)
\label{pf}
\ee
where $\tilde\varrho$ and $\tilde p$ are the energy density and pressure of the fluid, while $\tilde u_\mu$ is its unit velocity vector field (in what follows, I will always use tildes for the energy-momentum quantities of the $F(R)$ theories, to distinguish them from the corresponding GR quantities).
Conditions (\ref{Isrnew}) immediately imply then
\bea
\left. \tilde p\right|_{\Sigma}=0 , \label{p}\\
\left. \tilde\varrho \, n^\mu\tilde u_\mu \right|_{\Sigma}=0 . \label{pp}
\eea
The first of this determines the feasible matching hypersurfaces for perfect fluids, while the second informs us that the fluid must be either tangent to the matching hypersurface or with vanishing energy density there; they are both reminiscent of the GR case. 

\section{Discussion}
\label{sec:discussion}
Let me finally discuss the consequences of the above results, in particular in relation to whether or not matched solutions in GR are solutions of the extended $F(R)$ theories. The answer is generally no, as I am going to argue. I restrict myself now to the classical 4-dimensional situation: $n=4$. 

Consider any vacuum solution in GR (with a possible cosmological constant $\Lambda$). This is always a vacuum solution of (\ref{fe}) too provided $F(0)=-2\Lambda$. The metric is such that 
$$R^{+}_{\mu\nu}=\Lambda g^{+}_{\mu\nu}\quad \Longrightarrow \quad R^{+}=4\Lambda
$$ 
hence
$$
\nabla_{\mu}R^{+}=0.
$$
Thus, if it is to be matched to an interior solution, the junction conditions (\ref{Rcont}) and (\ref{DRcont}) imply that, on $\Sigma$, 
$$
R^{-}|_{\Sigma}=4\Lambda , \quad \quad \nabla_{\mu}R^{-}|_{\Sigma}=0.
$$
This happens to be generically incompatible with an interior metric that matches the vacuum solution in GR. The reason is that ---letting aside the fact that the matching hypersurface might not be the same as in GR--- the scalar curvature of the interior solution would not be constant on $\S$, even less with vanishing derivative, unless in very particular situations. 

Imagine, for instance, that the interior is described by a perfect fluid {\em in GR}:
\be
T_{\mu\nu}^{GR}=\varrho u_\mu u_\nu +p(g_{\mu\nu}+u_\mu u_\nu) \label{pfGR}
\ee
where $\varrho$ and $p$ are the energy density and isotropic pressure of the GR perfect fluid, respectively. 
Then, the scalar curvature in the interior region will be given by 
$$
R^{-}=-\kappa T^{GR}{}^\mu{}_\mu +4\Lambda= \kappa (\varrho -3p)+4\Lambda
$$
It follows that only the cases with 
$$
(\varrho -3p)|_{\Sigma}=0
$$ 
may survive as matched solutions in the extended theories. Notice that then $\varrho|_{\Sigma}=0$, for the matching hypersurface satisfies $p|_{\Sigma}=0$ in GR. This already discards many global solutions because matched solutions in GR may certainly have a jump in the energy density; as a matter of fact, this is what one wishes to model in some situations, such as constant density perfect fluids. 

The surviving GR solutions are even more scarce when this is supplemented  with the last condition 
$$
\nabla_{\mu}(\varrho -3p)|_{\Sigma}=0
$$
which will certainly rule out many of the few remaining cases.
Therefore, most GR solutions containing the given vacuum solution as the exterior are not solutions of the generalized $F(R)$ theory.

\section{Explicit examples} 
I present two illustrative and important examples. The first one considers the general case of static and spherically symmetric stars, the second deals with dynamical cases such as collapsing stars, eventually producing black holes.

\subsection{Static spherically symmetric perfect-fluid stars}
Consider any static spherically symmetric line-element
$$
ds^{2-}= -A^{2}(r) c^2 dt^{2}+\frac{dr^{2}}{1-2m(r)/r}+r^{2}d\Omega^{2}
$$
where $r$ is the area coordinate (round spheres at constant $t$ and $r$ have an area of $4\pi r^{2}$) and $d\Omega^{2}$ is the metric on the unit round sphere, $m(r)$ is the so-called mass function \cite{MTW}, and $A(r)$ is a function of $r$ that solves the Einstein's field equations for a perfect-fluid energy-momentum tensor (\ref{pfGR}) with velocity vector $u_{\mu}=-A c\delta_{\mu}^{t}$. A particular example is given by the interior Schwarzschild constant-density solution, see \cite{Exact} or \cite{W}, but there are many others. 

Assume that any such solution has been matched to the exterior Schwarzschild solution (I set $\Lambda =0$ now for simplicity). The matching hypersurface for these cases is given by the constant value of the area coordinate $r=r_{0}$ such that 
$$
p(r_{0})=0
$$
and then the mass $M$ of the exterior Schwarzschild part is
$$
M=\frac{8\pi}{c^2\kappa} m(r_{0}) .
$$
Eq.(\ref{Rcont}) implies then $\varrho(r_{0})=0$, and this by itself forbids many important cases, including the mentioned constant-density solution for the interior region. However, even for those cases with a GR equation of state $p=p(\varrho)$ such that $\varrho(r_{0})=0$ holds, Eq.(\ref{DRcont}) still requires that 
$$
\frac{d(\varrho-3p)}{dr }(r_{0})=0 \hspace{2mm} \mbox{in GR}
$$
and this of course gets rid of many sensible GR solutions. Actually, apart from the case of pure radiation ($p=\varrho/3$) in GR, only those interiors with 
$$
\frac{dp}{dr}(r_{0})=\frac{d\varrho}{dr}(r_{0})=0 \hspace{2mm} \mbox{in GR}
$$ 
survive.

Of course, there are many static spherically symmetric interior solutions matching the Schwarzschild exterior in $F(R)$ theories of gravity. The previous analysis just proves that the matched perfect-fluid solutions in GR will not be among them generically.

\subsection{Collapsing stars}
As for the second example, I consider dynamical situations. The simplest case is given by a Robertson-Walker (RW) interior region
\be
ds^{2-}= -c^{2}dt^{2}+a^{2}(t) d\Omega^{2}_{k} \label{RW}
\ee
where $a(t)$ is the scale factor and $d\Omega^{2}_{k}$ is the complete Riemannian 3-dimensional metric of constant curvature $k=\pm 1,0$. I denote by $\varrho$ and $p$ the GR energy density and pressure of the  fluid (\ref{pfGR}) for the RW geometry, that is to say, \cite{MTW,W}
$$
\kappa \varrho = 3\frac{\dot a^{2}+k}{a^{2}}, \hspace{2cm} \kappa (\varrho+3p) =-6\frac{\ddot a}{a}
$$
where dots denote derivatives with respect to $ct$. An elementary calculation using the field equations (\ref{fe}) proves that the very same metric is also a solution of the $F(R)$ field equations for an energy-momentum tensor of a perfect fluid (\ref{pf}) with energy density and pressure given by
\bea
\tilde\varrho =\frac{F'(R)}{2}(\varrho + 3p) +\frac{F(R)}{2\kappa} +F''(R) 3\frac{\dot a}{a} (\dot\varrho -3\dot p), \label{RWdens} \hspace{2cm} \\
\tilde p = \frac{F'(R)}{2}(\varrho  -p)-\frac{F(R)}{2\kappa}-F''(R)\{\ddot\varrho-3\ddot p +\frac{2\dot a}{a}(\dot\varrho -3\dot p)\}-F'''(R) (\dot\varrho -3\dot p)^2 \label{RWpre}
\eea
where in these formulae
$$
R = \kappa (\varrho -3p) = 6\left(\frac{\dot a^{2}+k}{a^{2}}+ \frac{\ddot a}{a}\right) \, .
$$

As is well-known, the RW metric (\ref{RW}) matches the Schwarzschild vacuum solution in GR whenever $p=0$, see e.g. \cite{MMV}, so the fluid becomes dust and $\varrho(t) a^{3}(t) =C=$const. The matching hypersurface is co-moving with the RW fluid flow. Actually, any such co-moving hypersurface works, and then the exterior mass $M$ gets determined accordingly. The Oppenheimer-Snyder collapse \cite{OS} to form a black hole, see e.g. \cite{W,MTW}, is included here for the closed case $k=1$. It is important to remark that the very same matching describes the Einstein-Straus model \cite{ES} of a vacuole in an expanding universe, as they are complementary matchings in the sense described in \cite{FST}, see \cite{MMV}. 

In contradistinction, it is impossible that such RW dust solution matches the Schwarzschild vacuum for general $F(R)$. To prove it, observe that the condition (\ref{Rcont}) would require 
$$
\varrho |_{\Sigma} =0
$$
but, given that $\varrho(t)$ depends only on $t$ and that $\Sigma$ is timelike, this would lead inevitably to 
$$
\varrho(t)=0 .
$$
Hence, the Oppenheimer-Snyder collapse or the Einstein-Straus vacuole in GR are no longer solutions in $F(R)$ gravity. The impossibility of the latter (and other RW cavities) in $F(R)$ gravity has been recently obtained in \cite{CDGN}.

One can further prove that actually no RW space-time can be matched to the Schwarzschild solution in $F(R)$ theories with $F''(R)\neq 0$ (again, I am assuming $F(0)=0=\Lambda$ for simplicity). To that end, observe that Eq.(\ref{Rcont}) implies that $R^{-}|_{\Sigma}=0$ and therefore 
$$
(\varrho-3p)|_{\Sigma}=0.
$$
Given that the matching hypersurface is timelike, this implies 
\be
\varrho -3p =0=R^{-} \label{R=0}
\ee
everywhere. Expressions (\ref{RWdens},\ref{RWpre}) show then that the energy-momentum tensor of the RW geometry in the $F(R)$ theory is described by a comoving perfect fluid with energy density $\tilde\varrho$ and pressure $\tilde p$ given by
$$
\tilde\varrho = F'(0) \varrho , \hspace{2cm} \tilde p =F'(0) p
$$
so that the fluid has the same radiation equation of state as in GR: $\tilde\varrho - 3 \tilde p=0$. But then the matching condition (\ref{p}) implies $\tilde p(t) =p(t)=0$, and {\it a fortiori} $\varrho (t)=0=\tilde\varrho (t)$ too because of (\ref{R=0}). 

One may wonder if these results depend crucially on the assumption of vacuum on the exterior, or on the particular nature of the RW metric. The answer is no once again, as similar conclusions follow for general radiating stars with spherical symmetry. To describe the exterior of such a star one can use the radiating Vaidya metric \cite{V,Exact}
$$
ds^{+2}=-\left(1-\frac{2m(u)}{r}\right)du^{2} -2du dr +r^{2} d\Omega^{2}
$$
where $m(u)$ is the mass function and  $u$ is null retarded time. This is a solution of Einstein's field equations for null incoherent radiation (it reduces to Schwarzschild for $m(u)=M=$const.)
$$
G^{+}_{\mu\nu}= R^{+}_{\mu\nu}=\frac{2}{r^{2}}\frac{dm}{du} \ell_{\mu}\ell_{\nu}, \hspace{3mm} \ell_{\mu}=-u_{,\mu}, \hspace{3mm} \ell_{\mu}\ell^{\mu}=0
$$
and it is easy to check that this is also a solution of (\ref{fe}) with $F(0)=0$ and
$$
\kappa T^{+}_{\mu\nu}=F'(0)\frac{2}{r^{2}}\frac{dm}{du} \ell_{\mu}\ell_{\nu}.
$$
It is known that a very large class of spherically symmetric metrics match the Vaidya solution \cite{FJLS}. The only requirement is the existence of a timelike hypersurface $\Sigma$ complying with conditions (\ref{Isr}), which essentially amounts to finding a hypersurface such that $n^{\mu}T^{-}_{\mu\nu}$ is null. The majority of these spherically symmetric interiors will have $R^{-}\neq 0$ and/or $\nabla R^{-} \neq 0$ so that they will no longer match the Vaidya solutions in $F(R)$ theories. 

Actually, in GR the RW metrics (\ref{RW}) always match the Vaidya solution and the matching hypersurface is timelike (in general not co-moving) whenever the dominant energy condition holds: $(p/\varrho)^{2}\leq 1$ \cite{FST}. The mass function gets then determined accordingly. The majority of these RW-Vaidya matched models are not solutions in $F(R)$ theories, because given that $R^{+}=0$, (\ref{Rcont}) implies $R^{-}=0$ and then the argument proceeds as above proving that $\varrho =3p$, in which case 
$$
\varrho(t) a^{4}(t)=\mbox{const.}
$$ 
Thus, only the pure radiation RW metric matched to Vaidya is also a global solution for arbitrary $F(R)$, keeping the matching hypersurface and the same equation of state in the $F(R)$ interior: $\tilde\varrho = 3 \tilde p$. This is certainly a very meager surviving set. Still, it is an explicit example of a {\em matched} solution which satisfies the GR field equations as well as the field equations of $F(R)$ theories.

\section*{Acknowledgements}
I thank M. Bouhmadi-L\'opez, S. Capozziello, T. Clifton, M. Mars, D. S\'aez-G\'omez, and R. Vera for comments and basic information. Supported by grants
FIS2010-15492 (MICINN), GIU06/37 (UPV/EHU), P09-FQM-4496 (J. Andaluc\'{\i}a---FEDER) and UFI 11/55 (UPV/EHU).

\section*{Appendix}
Let $(M^{\pm},g^{\pm})$ be two smooth $n$-dimensional spacetimes whose respective metrics are $g^{\pm}$. Assume that there are corresponding timelike hypersurfaces $\Sigma^{\pm}\subset M^{\pm}$ which bound the regions $V^{\pm}\subset M^{\pm}$ on each $\pm$-side to be matched. These two hypersurfaces are to be identified in the final glued spacetime, so that they must be diffeomorphic. The glued manifold is defined as the disjoint union of $V^{+}$ and $V^{-}$ with diffeomorphically related
points of $\Sigma^{+}$ and $\Sigma^{-}$ identified. Henceforth, this identified
hypersurfce will be denoted simply by $\Sigma$. An indispensable requirement to build a well-defined space-time ---with (at least) continuous metric--- is that the first fundamental forms $h^{\pm}$ of $\Sigma$ calculated on both sides agree because then
there is a metric extension $g$ defined on the entire manifold that coincides with
$g^{\pm}$ in the respective $V^{\pm}$ and is continuous \cite{CD,MS}.

In practice, one is given two spacetimes  and thus two sets of local coordinates $\{x^\mu_{\pm}\}$ {\em with no relation whatsoever} \cite{I}. Hence, one has two parametric expressions $x_{\pm}^\mu =x_{\pm}^\mu (\xi^a)$ of $\Sigma$, one for each imbedding into each of $M^\pm$, where $\{\xi^a\}$ are intrinsic local coordinates
for $\Sigma$ ($\mu,\nu,\dots  = 0,1,\dots,n-1; \, \, a,b,\dots =1,\dots , n-1$). The agreement of the two $(\pm)$-first fundamental forms amounts to the equalities on $\Sigma$
$$
h^+_{ab} = h_{ab}^- ,\hspace{1cm}
h^\pm_{ab} \equiv g_{\mu\nu}^\pm (x(\xi))\frac{\partial x_\pm^\mu}{\partial \xi^a}\frac{\partial x_\pm^\nu}{\partial \xi^b} \, .
$$
Denote by $n^\pm_\mu$ two unit normals to $\Sigma$ (one for each side). They are fixed up to a sign by the conditions
$$
n^\pm_\mu \frac{\partial x_\pm^\mu}{\partial \xi^a}=0, \quad n^\pm_\mu n^{\pm\mu}=1
$$
and one must choose one of them (say $n^-_\mu$) pointing outwards from $V^-$ and the other ($n^+_\mu$) pointing towards $V^+$. The two bases on the tangent spaces 
$$
\{n^{+\mu},\frac{\partial x_+^\mu}{\partial \xi^a}\} \quad \quad \{n^{-\mu},\frac{\partial x_-^\mu}{\partial \xi^a}\}
$$ are then identified, so that one can drop the $\pm$. The space-time version of the now unique first fundamental form is described by the projector to $\Sigma$
$$
h_{\mu\nu}=g_{\mu\nu}-n_\mu n_\nu .
$$
Notice that 
$$
h_{\mu\nu}\frac{\partial x^\mu}{\partial \xi^a} \frac{\partial x^\nu}{\partial \xi^b} =h_{ab}.
$$

At this stage, the Einstein field equations of GR are well-defined in the distributional sense, because one can easily prove \cite{MS,CD} that the Riemann tensor distribution (distributions are distinguished by an underline) takes the explicit expression
\be
\underline{R}^\alpha{}_{\beta\mu\nu}=(1-\underline{\theta}) R^{-\alpha}{}_{\beta\mu\nu}+\underline{\theta} R^{+\alpha}{}_{\beta\mu\nu}+\underline{\delta}^\Sigma H^\alpha{}_{\beta\mu\nu} \label{riedist}
\ee
where $R^{\pm\alpha}{}_{\beta\mu\nu}$ are the Riemann tensors of $V^\pm$ respectively, $\underline\theta$ is the distribution associated to the function that equals 1 on $V^+$ and vanishes on $V^-$, and $\underline{\delta}^\Sigma$ is a scalar distribution (a Dirac delta) with support on $\Sigma$ acting on any test function $Y$ by returning the value of the integral of this function on $\Sigma$:
$$
\left<\underline{\delta}^\Sigma ,Y\right> =\int_\S Y \, .
$$
It should be observed that
$$
\nabla_\mu\,   \underline{\theta} = n_\mu\,  \underline{\delta}^\Sigma \, .
$$
$H^\alpha{}_{\beta\mu\nu}$ in (\ref{riedist}) is called the singular part of the Riemann tensor distribution for obvious reasons, and should only be retained in idealized cases such as braneworlds or thin shells of matter. It has an explicit expression 
\be
H_{\alpha\beta\lambda\mu} = - n_{\alpha}\left[K_{\beta\mu}\right]n_{\lambda}+n_{\alpha}\left[K_{\beta\lambda}\right]n_{\mu}-n_{\beta}\left[K_{\alpha\lambda}\right]n_{\mu}+n_{\beta}\left[K_{\alpha\mu}\right]n_{\lambda} \label{HRie}
\ee
in terms of the jump of the second fundamental form across $\Sigma$:\footnote{Here the standard notation for discontinuities is used, so that for any function $f$ with definite limits on both sides of $\Sigma$, one sets for all $p\in \Sigma$: $\left[f\right](p)\equiv \lim_{x\rightarrow p}f^+(x) -\lim_{x\rightarrow p} f^-(x)$, $f^\pm$ being the restrictions of $f$ to $V^\pm$ respectively.}
$$
\left[K_{\beta\mu}\right]=K^+_{\beta\mu}- K^-_{\beta\mu}, \hspace{1cm} K^\pm_{\beta\mu}=h^\rho{}_{\beta}h^\sigma_\mu \nabla^\pm_\rho n_\sigma \, .
$$
Observe that each second fundamental form is symmetric and orthogonal to $n^\mu$, thus only the $n(n-1)/2$ components  tangent to $\Sigma$ are non-zero. A convenient formula for these components is 
$$
K^\pm_{ab} \equiv -n^\pm_\mu \left(\frac{\partial^2 x^\mu_\pm}{\partial\xi^a\partial\xi^b}Ê+\Gamma^{\pm \mu}_{\rho\sigma}\frac{\partial x_\pm^\rho}{\partial \xi^a} \frac{\partial x_\pm^\sigma}{\partial \xi^b}\right) \, .
$$
From (\ref{HRie}) the singular parts of the Ricci tensor $\underline{R}_{\beta\mu}$ and scalar curvature $\underline R$ distributions are easily computed to be, respectively
\be
H^\rho{}_{\beta\rho\mu}\equiv H_{\beta\mu}=-\left[K_{\beta\mu}\right] -\left[K^\rho{}_{\rho}\right] n_\beta n_\mu, \hspace{1cm} H^\rho{}_{\rho}\equiv H=-2\left[K^\mu{}_{\mu}\right] \label{HRic}
\ee
from where the singular part ${\cal G}_{\beta\mu}$ of the Einstein tensor distribution 
$$
\underline{G}_{\beta\mu}\equiv \underline{R}_{\beta\mu}-\frac{1}{2}g_{\beta\mu}\underline R = G^{+}_{\beta\mu}\underline\theta +G^{-}_{\beta\mu}(1-\underline\theta)+{\cal G}_{\beta\mu}\underline{\delta}^\Sigma
$$
follows \cite{I}
\be
{\cal G}_{\beta\mu} = -\left[K_{\beta\mu}\right]+h_{\beta\mu}\left[K^\rho{}_{\rho}\right] , \hspace{1cm} n^\mu {\cal G}_{\beta\mu} =0 \, . \label{HEi}
\ee
In GR, via the Einstein field equation, this provides the singular part $\tau_{\mu\nu}$ of the energy-momentum tensor distribution 
\be
\kappa \tau_{\beta\mu} = -\left[K_{\beta\mu}\right]+h_{\beta\mu}\left[K^\rho{}_{\rho}\right] , \hspace{1cm} n^\mu \tau_{\beta\mu} =0 \quad \mbox{(only in GR)} \label{HEin}
\ee
where $\kappa$ is the gravitational coupling constant: this is known as the Israel formula. Observe that only the tangent components ${\cal G}_{ab}$ and $\tau_{ab}$ are non-identically zero.

As a general result, the Bianchi identity $\nabla_{\rho}\underline{R}^\alpha{}_{\beta\mu\nu}+\nabla_{\mu}\underline{R}^\alpha{}_{\beta\nu\rho}+\nabla_{\nu}\underline{R}^\alpha{}_{\beta\rho\mu}=0$ holds in the distributional sense \cite{MS}, from where one deduces $\nabla^\beta \underline{G}_{\beta\mu}=0$ for the Einstein tensor distribution. A standard calculation with distributions then leads to
\bean
0= \nabla^\beta \underline{G}_{\beta\mu} = n^\beta \left[G_{\beta\mu}\right] \underline{\delta}^\Sigma +\nabla^\beta \left({\cal G}_{\beta\mu}\underline{\delta}^\Sigma \right)\\
=\underline{\delta}^\Sigma\left(n^\beta \left[G_{\beta\mu}\right] +\overline\nabla^\beta {\cal G}_{\beta\mu}-\frac{1}{2}n_\mu {\cal G}^{\rho\sigma}(K^+_{\rho\sigma}+K^-_{\rho\sigma})\right)
\eean
where $\overline\nabla$ denotes the intrinsic covariant derivative within $\Sigma$ associated to the first fundamental form. This implies the following relations (valid in general and well known in GR)
\bea
(K^+_{\rho\sigma}+K^-_{\rho\sigma}){\cal G}^{\rho\sigma} = 2n^\beta n^\mu \left[ G_{\beta\mu}\right]=2n^\beta n^\mu \left[ R_{\beta\mu}\right]-[R], \label{1}\\
\overline\nabla^\beta {\cal G}_{\beta\mu}=-n^\rho h^\sigma{}_\mu \left[ G_{\rho\sigma}\right]=-n^\rho h^\sigma{}_\mu \left[ R_{\rho\sigma}\right] \label{2} .
\eea
In GR, but not in general $F(R)$ theories, these can be trivially rewritten in terms of the energy-momentum tensor $T_{\mu\nu}$ and its singular distributional part $\tau_{\mu\nu}$ via the Einstein field equations.

An important remark is that these equations can also be obtained by using part of the Gauss and Codazzi equations for $\Sigma$ on both sides, specifically \cite{I}
\bea
R^{\pm}-2R^{\pm}_{\mu\nu}n^\mu n^\nu = {\cal R} -(K^{\pm\rho}{}_\rho)^2+K^{\pm}_{\mu\nu}K^{\pm\mu\nu},\label{gauss}\\
n^{\mu}R^{\pm}_{\mu\rho}h^{\rho}{}_{\nu}=\overline\nabla^{\mu}K^{\pm}_{\mu\nu}-\overline\nabla_{\nu}K^{\pm\rho}{}_{\rho} \label{coda}
\eea
where ${\cal R}$ is the scalar curvature of the first fundamental form of $\Sigma$.

Consider now the general case of $F(R)$-theories. The field equations read \cite{CF}
\be
F'(R)R_{\mu\nu}-\frac{1}{2}F(R)g_{\mu\nu}-\nabla_{\mu}\nabla_{\nu}F'(R)+g_{\mu\nu}\nabla_{\rho}\nabla^{\rho} F'(R) =\kappa T_{\mu\nu} \label{fe}
\ee
where primes denote derivatives with respect to $R$.
Alternatively, these equations can be written as follows
\bean
F'(R)R_{\mu\nu}-\frac{1}{2}F(R)g_{\mu\nu}-F''(R)\left( \nabla_\mu \nabla_\nu R -g_{\mu\nu} \nabla_\rho\nabla^\rho R\right)\\
-F'''(R)\left(\nabla_\mu R \nabla_\nu R -g_{\mu\nu} \nabla_\rho R \nabla^\rho R \right)=\kappa T_{\mu\nu} 
\eean
where we can see that covariant derivatives of $R$ (up to the second order) are needed to compute (\ref{fe}), and the only way to do that for a distribution $\underline R$ with a singular part or for a possibly discontinuous function $R$ is by the use of distribution theory.  One checks that, unless $F''(R)=0$ ---which corresponds to standard GR---, there are terms of type $\nabla_\mu \nabla_\nu \underline R$ and $\nabla_\mu \underline R \nabla_\nu \underline R$ which involve singular terms of type 
$$F(R)g_{\mu\nu}, \quad \quad F'(R)H_{\mu\nu}, \quad \quad F''(\underline{R})\nabla_\mu\nabla_\nu (H\underline\delta^{\Sigma}), \quad \quad
F''(\underline{R})\nabla_\mu ([R] n_\nu \underline\delta^{\Sigma})
$$ 
and also 
$$
F'''(\underline{R})\left(\nabla_{\mu}(H\underline\delta^{\Sigma})+ [R] n_{\mu} \underline\delta^{\Sigma}\right)\left(\nabla_{\nu}(H\underline\delta^{\Sigma})+ [R] n_{\nu} \underline\delta^{\Sigma}\right)
$$ 
containing products that are not allowed distributions\footnote{There are some recent results to treat the problem of multiplying distributions, see e.g. \cite{SV}, but still there is no unequivocal answer to this problem.} and that can {\em never} cancel with each other in (\ref{fe}). In order for the equations (\ref{fe}) to make sense even in the distributional sense, a quick analysis implies that the singular part of the scalar curvature distribution must vanish $H=0$, and from the second in (\ref{HRic}) this entails inevitably
$$
\left[K^\mu{}_\mu\right] =0 \, .
$$
Similarly, the requirement
$$
 \left[R\right]=0 
$$
is unavoidable {\em unless $F'''(R)=0$}. There arises an exceptional case for quadratic theories where the discontinuity of the scalar curvature can be non-zero and still the field equations make sense in the distributional sense. I analyze this case separately.

\subsection*{The generic case $F'''(R)\neq 0$.}
Taking (\ref{Kcont}) and (\ref{Rcont}) into account, the only remaining singular parts in the field equations (\ref{fe}) are those coming from the Ricci tensor distribution ---the first in (\ref{HRic})--- and the singular part of
$$
\nabla_{\nu}\nabla_{\mu}\underline R =(1-\underline{\theta}) \nabla_{\nu}\nabla_{\mu}R^{-}+\underline{\theta}\, \nabla_{\nu}\nabla_{\mu}R^{+}+\left[\nabla_{\mu}R\right] n_{\nu}\underline\delta^{\Sigma} \, .
$$
Taking (\ref{Rcont}) into account, the discontinuity $\left[\nabla_{\mu}R\right]$ is easily computed \cite{MS} to give
\be
\left[\nabla_{\mu}R\right] =a n_\mu , \hspace{1cm} a = n^\mu \left[\nabla_{\mu}R\right] \label{DRdisc}
\ee
where $a$ is a function defined on $\Sigma$ that represents the jump in the (normal) derivative $\nabla R$. Thus, the singular part of the field equations becomes (compare with (\ref{HEin}), and see \cite{DSS})
$$
\kappa \tau_{\mu\nu}= -F'(R_\Sigma) \left[K_{\mu\nu}\right] + F''(R_\Sigma) n^\rho \left[\nabla_{\rho}R\right] h_{\mu\nu},
\hspace{1cm} n^\mu \tau_{\mu\nu} =0
$$
where $R_\Sigma$ denotes the value of $R$ at $\Sigma$.

Using (\ref{Kcont}) and (\ref{Rcont}), the relations (\ref{1}-\ref{2}) become now
\bean
(K^+_{\rho\sigma}+K^-_{\rho\sigma}){\cal G}^{\rho\sigma}= (K^+_{\rho\sigma}+K^-_{\rho\sigma})H^{\rho\sigma} = 2n^\beta n^\mu \left[ R_{\beta\mu}\right],\\
\overline\nabla^\beta {\cal G}_{\beta\mu}=\overline\nabla^\beta H_{\beta\mu}=-n^\rho h^\sigma{}_\mu \left[ R_{\rho\sigma}\right] \, .
\eean
Even though these are different from the GR case, one can still wonder about the versions of (\ref{1}-\ref{2}) that involve the energy-momentum quantities. To derive them, the divergence of (\ref{newIsr}) is
$$
-F'(R_\Sigma) n^\rho h^\sigma{}_\mu \left[ R_{\rho\sigma}\right]- F''(R_\Sigma)\left[K_{\rho\mu}\right]\overline\nabla^{\rho}R_{\Sigma}+ F''(R_\Sigma)\overline\nabla_\mu a +F'''(R_{\Sigma})a\overline\nabla_{\mu}R_{\Sigma}=\kappa \overline{\nabla} ^\nu \tau_{\nu\mu}
$$
while from (\ref{fe}) the following discontinuity can be computed
$$
\kappa n^\rho h^\sigma{}_\mu \left[ T_{\rho\sigma}\right] = F'(R_\Sigma) n^\rho h^\sigma{}_\mu \left[ R_{\rho\sigma}\right]-F''(R_\Sigma) n^\rho h^\sigma{}_\mu \left[ \nabla_\rho\nabla_\sigma R\right] -F'''(R_{\Sigma})a\overline\nabla_{\mu}R_{\Sigma} \, .
$$
A straightforward standard calculation \cite{MS} leads to
\bea
\left[\nabla_{\mu}\nabla_{\nu}R\right] = A\,  n_{\mu}n_{\nu} +n_\mu \left(\overline\nabla_\nu a -\left[K_{\rho\nu}\right]\overline\nabla^{\rho}R_{\Sigma}\right)\nonumber \\ 
+n_\nu \left(\overline\nabla_\mu a -\left[K_{\rho\mu}\right]\overline\nabla^{\rho}R_{\Sigma}\right)+ \frac{a}{2} (K^+_{\mu\nu}+K^-_{\mu\nu}),
\label{D2R}
\eea
where $A$ is a function on $\Sigma$ defined by 
$$
A\equiv n^\mu n^\nu \left[\nabla_{\mu}\nabla_{\nu}R\right].
$$
Combining then the previous three expressions one arrives at
$$
\overline\nabla^\beta \tau_{\beta\mu}=-n^\rho h^\sigma{}_\mu \left[ T_{\rho\sigma}\right]. 
$$
Computing the total normal discontinuity of (\ref{fe}) and using (\ref{D2R}) and (\ref{Kcont}) one also gets
$$
(K^+_{\rho\sigma}+K^-_{\rho\sigma})\tau^{\rho\sigma} = 2 n^\beta n^\mu \left[ T_{\beta\mu}\right] . 
$$

\subsection*{The exceptional case $F'''(R)= 0$.}
Assume that $F'''(R)=0$, or equivalently that
\be
F(R) =R-2\Lambda +\alpha R^2 \label{Fquadratic}
\ee
for some constants $\Lambda$ and $\alpha$. Then a discontinuous $R$, with $[R]\neq 0$ is allowed in principle, from a mathematical point of view. Now, instead of (\ref{DRdisc}) one has
\be
\left[\nabla_{\mu}R\right] =a n_\mu +\overline\nabla_\mu [R], \hspace{1cm} a = n^\mu \left[\nabla_{\mu}R\right]  \label{DRdisc2}
\ee
and instead of (\ref{D2R}) 
\bea
\left[\nabla_{\mu}\nabla_{\nu}R\right] = A\,  n_{\mu}n_{\nu} +n_\mu \left(\overline\nabla_\nu a -\left[K_{\rho\nu}\right](\nabla^{\rho}R)_{\Sigma}-K^{\S}_{\rho\nu}\overline\nabla^{\rho}[R]\right)\nonumber \\ 
+n_\nu \left(\overline\nabla_\mu a -\left[K_{\rho\mu}\right](\nabla^{\rho}R)_{\Sigma}-K^{\S}_{\rho\mu}\overline\nabla^{\rho}[R]\right)+ a K^\S_{\mu\nu}+\overline\nabla_{\mu}\overline\nabla_{\nu}[R] 
\label{D2R2}
\eea
where now
$$
R_{\S}=\frac{1}{2}(R^{+}+R^{-}), \quad \quad (\nabla^{\rho}R)_{\Sigma}=\frac{1}{2}\nabla^{\rho}(R^{+}+R^{-})|_{\S}, \quad \quad K^{\S}_{\mu\nu}=\frac{1}{2}(K^{+}_{\mu\nu}+K^{-}_{\mu\nu}).
$$

The second derivative of the Ricci tensor distribution has then a singular part of the form
$$
\nabla_\nu\left([R]n_\mu \underline\delta^{\Sigma}  \right) +[\nabla_\mu R]n_\nu \underline\delta^{\Sigma} .
$$
which, after a calculation with distributions using (\ref{DRdisc2}) can be shown to adopt the explicitly symmetric form
\be
 \underline\delta^{\Sigma}\left\{\left(a-[R]K^\rho{}_\rho \right)n_\mu n_\nu +\frac{1}{2}[R](K^+_{\mu\nu}+
 K^-_{\mu\nu}) +n_\mu\overline\nabla_\nu [R] +n_\nu \overline\nabla_\mu [R] \right\}+\underline\Delta_{\mu\nu}
\ee
where $\underline\Delta$ is a 2-covariant symmetric tensor distribution whose components are defined, acting on any test function $Y$, by
$$
\left<\underline\Delta_{\mu\nu},Y\right> = -\int_\Sigma [R]n_\mu n_\nu\,  n^\rho\nabla_\rho Y \, .
$$
It should be observed that this distribution has support on $\S$ but it is of `$\delta'$' type, and thus its product with objects defined exclusively within $\S$ is not defined unless extensions of those objects off $\S$ are considered. 

The singular part of the left-hand side of the field equation (\ref{fe}) reads then
\bea
-\{1+\alpha(R^++R^-)\}[K_{\mu\nu}] \underline\delta^{\Sigma}+2\alpha \underline\Omega_{\mu\nu} \nonumber \hspace{2cm}\\
-2\alpha \underline\delta^{\Sigma}\left\{-ah_{\mu\nu} +\frac{1}{2}[R](K^+_{\mu\nu}+ K^-_{\mu\nu}-2K^\rho{}_\rho n_\mu n_\nu) +n_\mu\overline\nabla_\nu [R] +n_\nu \overline\nabla_\mu [R]  \right\} \label{excep}
\eea
where $\underline\Omega$ is a 2-covariant symmetric tensor distribution given by 
$$
\underline \Omega_{\mu\nu} = g_{\mu\nu} \underline\Delta^\rho{}_{\rho}-\underline\Delta_{\mu\nu}
$$
so that, acting on any test function $Y$, one has
\be
\left<\underline\Omega_{\mu\nu},Y\right> = -\int_\Sigma [R]h_{\mu\nu}\,  n^\rho\nabla_\rho Y \, . \label{Omega}
\ee

From (\ref{excep}) and the field equations (\ref{fe}) one observes that the energy-momentum tensor distribution is now allowed to have a singular part with several terms
$$
\underline T_{\mu\nu}=T^+_{\mu\nu} \underline\theta +T^-_{\mu\nu} (1-\underline\theta)+\tau_{\mu\nu}\,  \underline\delta^{\Sigma} + (\tau_\mu n_\nu +\tau_\nu n_\mu)\underline\delta^{\Sigma} +\tau n_\mu n_\nu \underline\delta^{\Sigma}+ \underline{t}_{\mu\nu}
$$
where 
\bean
\kappa \underline t_{\mu\nu}=2\alpha \underline\Omega_{\mu\nu} , \hspace{4cm}
\\
\kappa \tau_{\mu\nu} =-\{1+\alpha(R^++R^-)\}[K_{\mu\nu}] 
+\alpha \left\{2ah_{\mu\nu}-[R](K^+_{\mu\nu}+ K^-_{\mu\nu})\right\}, \, \, n^\mu\tau_{\mu\nu}=0, \\
\kappa \tau _\mu =-2\alpha \overline\nabla_\mu [R], \quad \quad n^\mu \tau_\mu =0, \hspace{4cm}\\
\kappa \tau = 2\alpha [R] K^\rho{}_\rho. \hspace{4cm}
\eean

\subsection*{Absence of thins shells or branes}
In proper cases where the matching hypersurface intends to describe the boundary between a matter interior and a vacuum exterior, or other similar separations, only jumps in the matter content (in the energy density, for instance) are to be allowed. Thus, one has to require that the singular terms in the energy-momentum tensor distributions vanish. From (\ref{newIsr}) or from (\ref{tauexc}-\ref{t}) this leads to
\be
\left[K_{\beta\mu}\right] =0 . \label{junc}
\ee
and to
$$
F''(R_\S)[R]=0,Ê\quad \quad F''(R_\S)n^\rho[\nabla_\rho R]=0\, .
$$
It is easily proven that (\ref{junc}) is equivalent to the vanishing of the whole Riemann singular part $H^\alpha{}_{\beta\mu\nu}$ \cite{MS}. If this holds, one can further prove that
\begin{itemize}
\item a local coordinate system around $\Sigma$ can be constructed such the metric is $C^1$ \cite{MS}. These are called admissible coordinates \cite{L1}, but they are not usually computed nor, indeed, explicitly constructible in analytical manner.
\item the Riemann tensor has a discontinuity described, in general, by the formula
\be
\left[R_{\alpha\beta\lambda\mu} \right]= n_{\alpha}B_{\beta\mu}n_{\lambda}-n_{\alpha}B_{\beta\lambda}n_{\mu}+n_{\beta}B_{\alpha\lambda}n_{\mu}-n_{\beta}B_{\alpha\mu}n_{\lambda} \label{disc}
\ee
where $B_{\beta\lambda}$ is a symmetric tensor defined only on $\Sigma$ which can be chosen to be tangent to $\Sigma$, that is to say, $n^\beta B_{\beta\lambda}=0$. Thus, $B_{\beta\lambda}$ contains $n(n-1)/2$ independent components, which are the allowed independent discontinuities of the Riemann tensor across $\Sigma$.
\end{itemize}
One can easily derive from (\ref{disc}) (or alternatively from (\ref{1}-\ref{2})) that the discontinuity of the Einstein tensor always satisfies
\be
n^{\mu}\left[G_{\mu\nu}\right]=0 \hspace{3mm} \Longleftrightarrow \hspace{3mm} 
n^{\mu}\left[R_{\mu\nu}\right]=\frac{1}{2}n_{\nu}\left[R\right] \label{Isr}
\ee
so that its $n$ normal components must be continuous across $\Sigma$. Actually, the $n(n-1)/2$ independent discontinuities of the Riemann tensor can always be chosen in GR to be those given by the discontinuities of the tangent components of the Einstein tensor (and thus, in GR, of the energy-momentum tensor). This is different in $F(R)$ of theories with $F''(R_\S)\neq 0$, as then $[R]=0$ necessarily.


\begin{thebibliography}{99}
\bibitem{SF} T.P. Sotiriou and V. Faraoni, Rev. Mod. Phys. {\bf 82} 451 (2010)
\bibitem{CF} S. Capozziello and V. Faraoni, {\em Beyond Einstein Gravity} (Springer, New York, 2011)
\bibitem{NO} S. Nojiri and S.D. Odintsov, Phys. Rep. {\bf 505} 59 (2011)
\bibitem{CDGN} T. Clifton, P. Dunsby, R. Goswami, and A.M. Nzioki, Phys. Rev. D {\bf 87}, 063517 (2013)
\bibitem{DSS} N. Deruelle, M. Sasaki and Y. Sendouda, Prog. Theor. Phys. {\bf 119} 237 (2008)
\bibitem{L} A. Lichnerowicz, C. R. Acad. Sci. {\bf 273} 528 (1971)
\bibitem{T} A. H. Taub, J. Math. Phys. {\bf 21} 1423 (1979)
\bibitem{BD} A. Balcerzak and M.P. Dabrowski, Phys. Rev. D {\bf 84} 063529 (2011)
\bibitem{OS} J.R. Oppenheimer and H. Snyder, Phys. Rev. {\bf 56} 455 (1939)
\bibitem{FST} F. Fayos, J.M.M. Senovilla, and R. Torres, Phys. Rev. D {\bf 54}, 4862 (1996)
\bibitem{ES} A. Einstein and E.G. Straus,  Rev. Mod. Phys. {\bf 17} 120 (1945); erratum {\bf 18} 148 (1946)
\bibitem{MMV} M. Mars, F. C. Mena, and R. Vera, Phys. Rev. D {\bf 78} 084022 (2008)
\bibitem{ESK} A.L. Erickcek, T.L. Smith and M.Kamionkowski, Phys. Rev. D {\bf 74}, 121501(R) (2006)
\bibitem{HS} W. Hu and I. Sawicki, Phys. Rev. D {\bf 76} 064004 (2007)
\bibitem{V} P.V. Vaidya, Proc. Indian Acad. Sci. A {\bf 33} 264 (1951)
\bibitem{F} A.V. Frolov, Phys. Rev. Lett {\bf 101} 061103 (2008).
\bibitem{BSM} E. Barausse, T.P. Sotiriou, and J.C. Miller, Class. Quantum Grav. {\bf 25} 062001 (2008); {\it ibid.} 105008 (2008)
\bibitem{KM} T. Kobayashi and K. Maeda, Phys. Rev. D {\bf 78} 064019 (2008)
\bibitem{MS} M. Mars and J.M.M. Senovilla, Class. Quantum Grav. {\bf 10} 1865 (1993)
\bibitem{D} G. Darmois, M\'emorial des Sciences Math\'ematiques, Fascicule 25 (Gauthier-Villars, Paris, 1927)
\bibitem{I} W. Israel, Nuovo Cimento {\bf 44}, 1 (1966); erratum {\bf 48}, 463  (1967)
\bibitem{L1} A. Lichnerowicz, {\it Th\'eories Relativistes de la Gravitation et de l'Electromagn\'etisme} (Masson, Paris, 1955).
\bibitem{Exact} H. Stephani, D. Kramer, M.A.H. MacCallum, C. Hoenselaers and E. Herlt,
{\it Exact Solutions to Einstein's Field Equations Second Edition}
(Cambridge University Press, Cambridge, 2003)
\bibitem{MTW} C.W. Misner, K.S. Thorne, and J.A. Wheeler,  {\it Gravitation},
(W.H. Freeman and Company, New York , 1970).
\bibitem{W} S. Weinberg, {\it Gravitation and Cosmology}, (Wiley, N. York, 1972).
\bibitem{FJLS} F. Fayos, X. Ja\'en, E. Llanta, and J.M.M. Senovilla, Phys. Rev. D {\bf 45} 2732 (1992).
\bibitem{CD} C.J.S. Clarke and T. Dray, Class. Quantum Grav. {\bf 4},
265 (1987)
\bibitem{SV} R. Steinbauer and J.A. Vickers, Class. Quantum Grav. {\bf 23}, R91 (2006)

\end{thebibliography}
\end{document}